# On the Suitability of the 2 x 2 Games for Studying Reciprocal Cooperation and Kin Selection

James A. R. Marshall[*]

[*]*Department of Computer Science, University of Bristol, BS8 1UB, UK. Tel: 44 (0)117 9545394 Fax: 44 (0)117 9545208 Email: James.Marshall@bristol.ac.uk*

**Abstract**

**The 2 x 2 games, in particular the Prisoner's Dilemma, have been extensively used in studies into reciprocal cooperation and, to a lesser extent, kin selection. This paper examines the suitability of the 2 x 2 games for modelling the evolution of cooperation through reciprocation and kin selection. This examination is not restricted to the Prisoner's Dilemma, but includes the other non-trivial symmetric 2 x 2 games. We show that the popularity of the Prisoner's Dilemma for modelling social and biotic interaction is justified by its superiority according to these criteria. Indeed, the Prisoner's Dilemma is unique in providing the simplest support for reciprocal cooperation, and additive kin-selected altruism. However, care is still required in choosing the particular Prisoner's Dilemma payoff matrix to use. This paper reviews the impact of non-linear payoffs for the application of Hamilton's rule to typical altruistic interactions, and derives new results for cases in which the roles of potential altruist and beneficiary are separated. In doing so we find the same equilibrium condition holds in continuous games between relatives, and in discrete games with roles.**

**Keywords**
2 x 2 games, Prisoner's Dilemma, reciprocal cooperation, kin selection, altruism, evolution of cooperation, evolutionary game theory, synergistic effect, roles

**Introduction**

Two of the main mechanisms proposed to explain the evolution of cooperation between self-interested individuals are reciprocity (Trivers, 1971; Axelrod, 1984) and kin selection (Hamilton, 1964a, 1964b). Game theory has frequently provided a fundamental tool for investigations into these two mechanisms (for examples see Trivers (1971), Axelrod (1984, 1987), and Nowak & May (1993), among many others). Previously the candidate mechanisms have been investigated separately, however both mechanisms are expected under similar circumstances, and thus are likely to interact with each other (Trivers, 1971). Furthermore, it has been suggested that kin selected altruism could have formed an initial stage in the evolution of reciprocal cooperation (Axelrod & Hamilton, 1981). Only a few theoretical studies have investigated reciprocal cooperation and kin selection together (Wilson & Dugatkin, 1991; Dugatkin *et al*., 1994; Dugatkin, 1997; Marshall & Rowe, 2000; Marshall & Rowe, 2003a), again using game theory. This paper examines the applicability of the different 2 x 2 games for the modelling of reciprocal cooperation, and kin selection for altruism. In doing so, we both vindicate the Prisoner's Dilemma as the best choice for such models, and identify important caveats in its application.

**Reciprocal Cooperation**

Reciprocal cooperation was proposed by Trivers (1971) and Axelrod (1984) as a means of explaining the evolution of cooperative behaviour between selfish individuals. Repeated interaction between such individuals is expected to facilitate cooperation if the individuals are able to reciprocate cooperative and non-cooperative behaviour. Axelrod (1984) discussed a concept known as the "shadow of the future", based on Shubik's work (1970), which is the probability of an interaction between two individuals continuing in the future. Axelrod used the Iterated Prisoner's Dilemma as the basis of his work and showed that if the shadow of the future is sufficiently large then cooperation can be favoured, even though in the single-iteration Prisoner's Dilemma mutual defection is the natural outcome. This is known as direct reciprocity. While this paper restricts itself to direct reciprocity, indirect reciprocity has also been proposed, in which one individual bases the decision to cooperate or not cooperate with another individual on the level of cooperative behaviour that individual exhibits in its interactions with others (Nowak & Sigmund, 1998).

**Kin Selection**

An alternative explanation for cooperation, or altruism, among selfish individuals was proposed by Hamilton (1964a, 1964b) in the form of kin selection. Kin selection proposes that an altruistic trait under genetic control may spread through a population, if the benefits of its altruism are on average directed towards copies of itself in that population. This is captured in Hamilton's rule (Hamilton, 1964a), which states that an altruistic gene will be selected for when

$$\frac{c}{b} < r \,, \tag{1}$$

where $c$ is the fitness cost to the donor of the altruistic act, $b$ is the fitness benefit of the act to the recipient, and $r$ is the degree of relationship between the donor and recipient. The degree of relationship $r$ is functionally equivalent to the probability that the recipient contains a copy of the altruistic allele present in the donor, over and above the frequency of the allele in the population at large (Grafen, 1985).

While it is possible to have a zero, or even negative, cost in Hamilton's rule, such acts are not truly altruistic but may be termed "weak altruism" (Wilson, 1979). As kin selection theory was developed to help explain altruistic acts, this paper limits itself to kin selection for "strong altruism" (*ibid.*), in other words behaviour with a positive fitness cost associated with it. Thus, we are interested in altruistic interactions of the form shown in table 1, which presents a payoff matrix showing deviation from base fitness for the donation game played between two partners. Altruism (A) provides a positive fitness benefit $b$ to the recipient, while incurring a positive fitness cost $c$ to the donor. Selfishness (a) has no fitness cost and provides no fitness benefit. Typically a pair of altruists interacting should each have higher fitness than a pair of non-altruists, so the donation game can be characterised by the inequality $b > c > 0$.

|   | A   | a  |
|---|-----|----|
| A | b – c | –c |
| a | b   | 0  |

Table 1. Donation Game Payoff Matrix
*(note: payoffs to the row player are shown)*

**The 2 x 2 Games**

We now consider the 2 x 2 games and their support for modelling reciprocal cooperation and kin-selected altruism. The 2 x 2 games are a class of games in which there are two participants, each of whom has a choice of two possible actions. As such they are the simplest possible form of game. A taxonomy of the 2 x 2 games was provided by Rapoport & Guyer (1966). Subsequently the most interesting 2 x 2 games to study were identified by Rapoport (1967) as being the non-trivial symmetric games. Non-triviality in this definition means that there is no outcome that is preferred by both players. Symmetry simply means that the game payoff orderings are identical from both players' points of view. The games thus identified by Rapoport (1967) were the Prisoner's Dilemma, Chicken, the Battle of the Sexes, and Apology. Of these, the Prisoner's Dilemma is by far the best known and most widely studied.

**Strategies for the 2 x 2 Games**

It has become common practice to study iterated interaction in a 2 x 2 game such as the Prisoner's Dilemma. Allowing the same players to interact repeatedly gives scope for some interesting dynamics, including the establishment of reciprocally cooperative behaviour. The players of a game, whether they interact repeatedly with each other or not, require a strategy to determine how they play the game. A strategy specifies an action choice for a player, which may or may not be contingent on some history of interaction with the opponent. The simplest strategies for the Prisoner's Dilemma are those of unconditional defection or cooperation, and these strategies do not take account of iterated interaction. Such strategies can still give rise to interesting dynamics, for example in spatial games (Nowak & May, 1992). A slightly more sophisticated strategy is the one-dimensional strategy, which refers to the opponent's action on the previous interaction, then specifies a current action choice based on this. More sophisticated still is the two-dimensional strategy, in which the current action is specified by both the player's previous action and its opponent's previous action. There are of course a limitless number of strategy types and representations; for two examples of alternative approaches see Axelrod (1987) and Chess (1988).

As well as taking account of interaction history in different ways, strategies can also be deterministic or stochastic; deterministic strategies specify a single action choice for each eventuality, whereas stochastic strategies specify a probability distribution for different action choices in each eventuality. Stochastic strategies may also be thought of as continuous strategies, which deterministically specify an action as a continuous rather than binary value.

In this paper, the 2 x 2 games shall be discussed in the context of one of the simplest conditional strategies, the two-dimensional deterministic strategy. While the simplest conditional strategies for the Prisoner's Dilemma, the one-dimensional strategies, include the classic strategy for reciprocal cooperation, Tit-for-Tat (Axelrod, 1984), they exclude other interesting examples such as the two-dimensional strategy

Pavlov (Nowak & Sigmund, 1993). As shall be shown in this paper, however, the best motivation for studying the two-dimensional strategies is that they include two fundamental types of altruistic interaction, those in which both players simultaneously decide whether or not to behave altruistically, and those in which the roles of potential altruist and potential beneficiary are separated.

**Strategy Encoding and Kin Selection**

A necessary condition for kin selection to occur is that altruistic behaviour be under genetic control, indicating that for game theoretic studies into kin selection, strategies must have a genetic representation. One approach is to make the traditional assumption of the evolutionary game theorist, that the different strategies are true-breeding phenotypes. However, this approach, described by Grafen (1984) as the "phenotypic gambit", is vulnerable to arbitrary strategy-set restrictions when dealing with strategies of any more complexity than simple unconditional behaviours. The approach taken in this paper will instead consider strategies that are specified by the interaction of several loci, as follows. Two-dimensional strategies can be represented in a variety of different ways; the approach used here will be adapted from that of Mar & St. Denis (1994) for two-dimensional continuous strategies, in which a separate action is defined for each possible interaction history, as well as one action for the first encounter between two players. This strategy can thus be encoded using five loci, with each locus determining behaviour in a given situation and having two alleles, cooperate (C) and defect (D). The alleles for these strategy loci are labelled in table 2 below.

| Previous actions (own / opponent's) | Cooperation allele | Defection allele |
|---|---|---|
| Initial action | friendly | suspicious |
| C/C | constructive | destructive |
| C/D | forgiving | vengeful |
| D/C | merciful | exploitative |
| D/D | dovish | hawkish |

Table 2. Two-Dimensional Strategy Loci and Alleles

Given such a strategy encoding, we could enumerate the strategies produced by the different possible combinations of alleles and undertake a classical evolutionary game theory analysis using this strategy set. For our chosen strategy encoding, this would yield $2^5$ possible strategies. Alternatively, we can simplify our analysis by restricting it to one locus or a subset of loci, and characterising the selective pressures for the different alleles at each of these loci. In this context, kin-selected altruism may occur if an allele at a given locus of a strategy provides a benefit to other strategies in the population containing the same allele at the same locus, while incurring a cost to the individual that it is expressed in.

**Games 1 and 2: Apology and Battle of the Sexes**

We now begin our examination of the 2 x 2 games and their suitability for modelling reciprocal cooperation and kin selection.

|   | A | B |
|---|---|---|
| A | -1 | 2 |
| B | 1 | -2 |

Table 3. Apology Payoff Matrix (Rapoport, 1967)
*(note: payoffs to the row player are shown)*

|   | A | B |
|---|---|---|
| A | -1 | 1 |
| B | 2 | -2 |

Table 4. Battle of the Sexes Payoff Matrix (Rapoport, 1967)
*(note: payoffs to the row player are shown)*

Tables 3 and 4 above present the ordinal payoff matrices for Apology and Battle of the Sexes respectively. These games belong to a class of games labelled "pre-emption games" by Rapoport & Guyer (1966). In each game there is an incentive for each player to be the first to pre-empt by choosing B while their opponent chooses A, thus securing the largest available payoffs for both. Such an outcome is a pareto-equilibrium (Rapoport & Guyer, 1966), meaning that there is no other outcome of the game where both players receive higher payoffs. The difference between Apology and Battle of the Sexes lies in which player takes the largest payoff when pre-emption occurs; in Battle of the Sexes this is the pre-empting player, while in Apology it is the pre-empted player. However there is a clear danger of both players pre-empting simultaneously, resulting in the lowest available payoff for both, while if neither player pre-empts then each player receives the second lowest payoff.

Reciprocal cooperation in Apology and Battle of the Sexes, in other words equitable sharing of the maximum payoffs available, is complicated. This is because the maximum available payoffs in these games are only available when one player pre-empts the other, and in such a situation the payoff to one player is always higher than the payoff to the other. First, co-ordination is required for the players to arrive at one of the pareto-optimal outcomes. Simply to cooperate to achieve a large payoff for both players is non-trivial, even without equitable sharing, as the best strategy to play will depend on the frequency with which the two strategies are encountered. For this reason a fully mixed population playing the non-iterated version of the game will converge on a mixed strategy equilibrium. For the iterated version of the game, two players interacting with each other repeatedly need to take turns in pre-empting in order to equitably share payoffs. As we shall see later, simpler models of cooperation are possible in the 2 x 2 games. This suggests that Apology and Battle of the Sexes are less than ideal for studying reciprocal cooperation.

Furthermore, Apology and Battle of the Sexes do not prove suitable for modelling kin selection. Decomposing Apology or Battle of the Sexes into sums of fitness costs and benefits, as in table 1, is not possible without assuming a negative cost for altruism; in other words assuming that an altruistic act increases the actor's fitness instead of decreasing it (see appendix B for the proof). As already discussed, in modelling kin selection we are interested foremost in situations of "strong altruism" such as the donation game, where altruism results in a net loss of payoff.

**Game 3: Chicken**

|   | A | B |
|---|---|---|
| A | 1 | -1 |
| B | 2 | -2 |

Table 5. Chicken Payoff Matrix (Rapoport, 1967)
*(note: payoffs to the row player are shown)*

Table 5 above presents the ordinal payoff matrix for Chicken, also known as the Snowdrift game. Like Apology and Battle of the Sexes, Chicken is a pre-emption game, with two Pareto equilibria corresponding to pre-emption by one or the other player. Unlike those two games, however, in Chicken there is a strong incentive to be the first to pre-empt and receive an increased payoff at the expense of the opponent. Like Apology and Battle of the Sexes, there is a clear danger of both players pre-empting simultaneously, resulting in the lowest available payoff for both, however unlike those games, if neither player pre-empts then each player receives the second highest payoff.

Chicken is similar to the Prisoner's Dilemma, discussed later, in that mutual cooperation can be effectively achieved by co-ordinating on the AA outcome, resulting in the second highest payoff for both players. Like the Prisoner's Dilemma, this outcome is vulnerable to one of the players changing their action choice, thus increasing their payoff at the expense of their opponent. Unlike the Prisoner's Dilemma, however, once one player has exploited the other in this manner, that exploitation is an equilibrium. This indicates that in the iterated version of the game, reciprocation of exploitation may not be an effective strategy, suggesting that Chicken is not well suited to studying the evolution of reciprocal cooperation.

Chicken is also unsuitable for correctly modelling kin selected altruism according to our requirements. As with Apology and Battle of the Sexes, mapping the game's payoff matrix onto the payoff matrix for the donation game (table 1) is only possible if a negative cost for altruism is assumed (see appendix B for further details). This is the approach taken by Doebeli & Hauert (2005), in which altruists receive the same fitness benefit that they give to the recipient of their altruism. As the benefit exceeds the cost in that formulation, there is no cost for altruism. The benefit from two altruists interacting in the game arises from halving the costs of altruism that each must bear (Doebeli & Hauert, 2005, table 1). Recalling that we are interested in modelling strong altruism, in which altruism has a fitness cost, Chicken is shown to be unsuitable for that purpose.

**Game 4: The Prisoner's Dilemma**

|   | C | D |
|---|---|---|
| C | 3 (*R*) | 0 (*S*) |
| D | 5 (*T*) | 1 (*P*) |

Table 6. Prisoner's Dilemma Payoff Matrix (Axelrod, 1984)
*(note: payoffs to the row player are shown, payoff labels are in brackets)*

The traditional cardinal payoff matrix in table 6 above is for the Prisoner's Dilemma. The cardinal rather than ordinal payoff matrix is presented here, as the Prisoner's Dilemma typically has the payoff constraint $T + S < 2R$, in addition to the ordinal constraint $T > R > P > S$. Also, the game choices A and B have been replaced with the more familiar C and D (for cooperation and defection respectively). The Prisoner's Dilemma is well known for having a single deficient equilibrium, namely mutual defection (D) (Rapoport & Guyer, 1966), which can be escaped in favour of mutual cooperation when the game is iterated (Shubik, 1970; Trivers, 1971; Axelrod, 1984). As has been discussed at length by other authors, the Prisoner's Dilemma is perfect for the first of our modelling objectives, modelling reciprocal cooperation. However, the support for kin-selected altruism in the Prisoner's Dilemma requires more careful consideration.

The first question to answer is if the Prisoner's Dilemma payoff matrix (table 6) can be mapped onto the payoff matrix for the donation game (table 1). The answer is yes, but certain conditions must be fulfilled, as follows. The key point to note in the donation game (table 1) is that costs and benefits of altruism are additive. An individual paying the cost of altruism to an opponent, while simultaneously receiving the benefit of their opponent's altruism, simply has a net fitness change $b - c$. However, the Prisoner's Dilemma payoffs do not guarantee that net fitness change will be additive. Non-additive fitness effects pose a serious problem for the application of Hamilton's rule (equation 1), which will result in inaccurate predictions on under which circumstances kin-selected altruism will be favoured. Queller (1984, 1985) addressed this problem by studying a form of the donation game extended with a new parameter, the synergistic effect that occurs when two altruists interact with each other (table 7).

|   | A | a |
|---|---|---|
| A | $b - c + d$ | $-c$ |
| a | $b$ | 0 |

Table 7. Payoff Matrix for Donation Game with Synergistic Effect (Queller, 1984)
*(note: payoffs to the row player are shown)*

For any given Prisoner's Dilemma payoff matrix, we can analyse it in terms of table 7 to determine the sign and magnitude of the synergistic effect $d$. $d$ is the deviation from additivity of cost and benefit that occurs when both individuals behave altruistically. If $d < 0$ then an altruist interacting with another altruist pays an additional fitness cost, and we may call the payoff matrix *negatively non-additive*. If $d > 0$ then an altruist interacting with another receives an additional fitness boost, and we may call the matrix *positively non-additive*. If $d = 0$ then the fitness of an altruist is independent of their opponent's phenotype, and the payoff matrix is additive. The game may now be analysed in the usual way for evolutionary stable strategies (ESS), and the effect that different kinds of deviation from additivity, positive or negative, have on them. This analysis has already been undertaken for a generic 2 x 2 game (Grafen, 1979; Queller, 1984; Queller, 1985[1]), and here we recount these results with reference to the Prisoner's Dilemma.

Queller (1984) maps the synergistic donation game payoffs given in table 7 onto a generic 2 x 2 game thus:

$$-c = B - D, \qquad (2a)$$

$$b = C - D, \qquad (2b)$$

and

$$d = A - B - C + D, \qquad (2c)$$

where $A$, $B$, $C$ and $D$ are the payoffs for a generic 2 x 2 game labelled Hawk-Dove by Queller. Mapping these payoff labels onto the Prisoner's Dilemma payoff labels we have been using so far, we thus have a condition that tells us when a Prisoner's Dilemma payoff matrix can be mapped onto the donation game payoff matrix (table 1), namely when

$$R - T + P - S = 0. \qquad (3)$$

Note that when using the canonical Prisoner's Dilemma payoffs $T = 5$, $R = 3$, $P = 1$ and $S = 0$, the synergistic effect $d$ when two altruists interact is negative.

Let us now examine the selective pressures on two different subsets of loci in our two-dimensional strategies. These first of these loci are those involved in interactions in which both opponents use the same locus to specify their current action, namely the "friendly / suspicious", "constructive / destructive" and "dovish / hawkish" loci. Assume a population structure in which individuals interact with an opponent containing the same allele with a certain probability $r$, otherwise they interact with an opponent drawn at random from the entire population. Then the fitness of cooperation and defection alleles at these loci may be written respectively as

$$w_c = rR + (1-r)(f_c R + (1-f_c)S), \qquad (4)$$

and

$$w_d = rP + (1-r)(f_c T + (1-f_c)P), \qquad (5)$$

where $f_c$ is fraction of cooperation alleles at that locus in the whole population.

Selection will favour cooperation alleles over defection alleles when $w_c > w_d$. This inequality can be re-arranged to give us a relatedness threshold $r'$ above which selection will favour cooperation alleles. Equation (6) can also be derived easily from Queller's (1984) equation (5a) for the spread of an altruistic allele in terms of cost, benefit and deviation from additivity:

$$r' = \frac{f_c(R-T+P-S)+S-P}{f_c(R-T+P-S)+S-R}. \tag{6}$$

---

[1] However, beware the incorrect reporting of Grafen's (1979) results (equations 6 and 9) in Queller's (1984) equations 1 and 2.

If the deviation from additivity $d$ of the payoff matrix used is zero, then the response to selection will be accurately predicted by Hamilton's rule (equation 1) with the ratio of cost $c$ to benefit $b$ being given by

$$\frac{S-P}{S-R}. \tag{7}$$

(7) does not on first sight look like the ratio of fitness cost ($c = P - S$) to fitness benefit ($b = T - P$), however if $d = 0$ then $R - T + P - S = 0$, which gives us $R - S = T - P$ and we recover Hamilton's rule.

If, however, $d$ is non-zero, the relatedness threshold above which altruism is favoured will vary as a function of $f_c$. If $d$ is negative, altruism will be harder to select for than predicted by Hamilton's rule, and when $d$ is positive, altruism will be easier to select for. With reference to equation (6) we can see that with a non-zero $d$ the relatedness threshold $r'$ for altruism to be selected for will vary in the closed interval between (7) and

$$\frac{R-T}{P-T}. \tag{8}$$

Note that when $d < 0$ (8) has the greater value, and when $d > 0$ (7) does.

As shown by previous authors, the mixed equilibrium frequency of cooperators in the population is given by

$$f_c' = \frac{P-S-r(R-S)}{(1-r)(R-T+P-S)}, \tag{9}$$

with this equilibrium being unstable if $R - S + P - T > 0$ and stable otherwise (Grafen, 1979; Queller, 1984). From equation (6) we can see that the equilibrium will be mixed only if $r$ is in the open interval whose bounds are given by (7) and (8), otherwise either cooperation or defection alleles will become fixed in the population depending on the value of $r$. For illustration, figure (1) shows the variable relatedness threshold for a particular negatively non-linear payoff matrix, the canonical Prisoner's Dilemma payoffs $T = 5$, $R = 3$, $P = 1$ and $S = 0$. Figure (1) also shows how, given a particular population structure realising a coefficient of relatedness $r = \frac{5}{12}$, the population converges on a mixed strategy equilibrium, in which cooperation and defection alleles are maintained in the population in a ratio of 3 to 4.

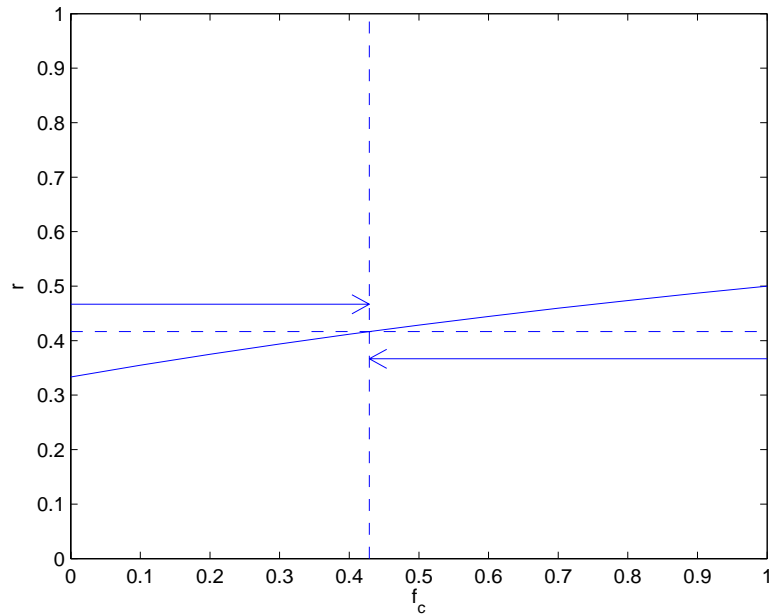

Figure 1. The Relatedness Threshold for Kin-Selected Altruism at a Single Locus to be Favoured with Negatively Non-Linear Payoffs
For an intermediate relatedness level a population will converge on a mixed equilibrium of cooperation and defection alleles ($T = 5$, $R = 3$, $P = 1$, $S = 0$)

Thus far we have reviewed existing results on discrete non-additive games played between relatives using a single locus, and shown how the Prisoner's Dilemma can be parameterised to give additive, positively non-additive or negatively non-additive payoff matrices. As well as kin selected altruism at a single locus, kin-selected altruism is possible in a 2 x 2 game when opponents each use a different locus to determine their current action choice. With deterministic two-dimensional strategies, as considered here, this occurs when each player chose a different action to the other on their last interaction together. At either of the loci controlling action choices of this kind, which are labelled the "merciful / exploitative" and "forgiving / vengeful" loci for this discussion, whether kin selection for altruism is favoured or not is determined by the relative inclusive fitness of cooperation and defection alleles, which in turn may be dependent on the frequency of cooperation alleles at another locus, as will be shown below. Thus, one player using a cooperation allele to determine their current action may provide an altruistic benefit to a relative containing the same cooperation allele, if that relative uses a defection allele at a different locus to determine their current action. This situation is illustrated in figure (2).

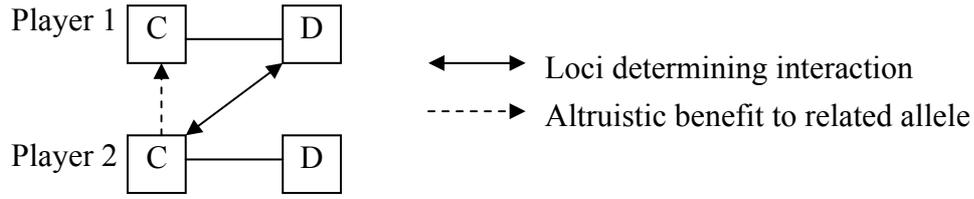

Figure 2. Altruistic Donation with Role Separation in the Prisoner's Dilemma
Player 1 is the recipient of player 2's altruism, which arises because player 2 cooperates (C) while player 1 defects (D). This reduces player 2's payoff and increases player 1's. Kin selection can favour the altruistic allele (C) in this case, as player 1 contains a copy of the same allele

To analyse selection at these loci we shall, as before, calculate the inclusive fitness of cooperation and defection alleles. Inclusive fitness is a weighted sum of an individual's reproductive fitness excluding that due to external effects, and the portion of the reproductive fitness of the relatives affected by that individual's behaviour, where the weights in question are the degrees of relatedness of those relatives to the individual (Hamilton, 1964a). Marshall & Rowe's (2000, 2003) calculation of these inclusive fitnesses for cooperation ($i_c$) and defection ($i_d$) alleles at the "forgiving / vengeful" locus of a two-dimensional strategy is as follows:

$$i_c = Rf_c + S(1-f_c) + r(Rf_c + T(1-f_c)), \qquad (10)$$

and

$$i_d = Tf_c + P(1-f_c) + r(Sf_c + P(1-f_c)), \qquad (11)$$

where $f_c$ represents the environmental frequency of cooperation alleles at the other locus and $r$ represents the relatedness for the allele at the locus under consideration (i.e. the probability that the individual being interacted with has the same allele over and above the population frequency of that allele (Grafen, 1985)). Note that the frequency of the altruist type $f_c$ is independent of the relatedness $r$ because the model presented here is an example of the situation, described by Queller (1984), in which the roles of potential altruist and potential beneficiary are separated. The reason for this in the current model can be understood by reference to figure 2, and is demonstrated in appendix A. Here the non-random determination of roles is due to the interaction history of the participants in the game.

Based on equations (10) and (11), the condition $i_c > i_d$ can be rearranged to give the threshold relatedness above which cooperation alleles are favoured by kin selection (from Marshall & Rowe (2003)):

$$r' = -\frac{f_c(R-T+P-S)+S-P}{f_c(R-T+P-S)+T-P}. \qquad (12)$$

Figure (3) below shows how, for negatively non-linear payoffs, the degree of relatedness $r'$ required for kin-selected altruism to be favoured varies as a function of the frequency $f_c$ of cooperative alleles at the other locus.

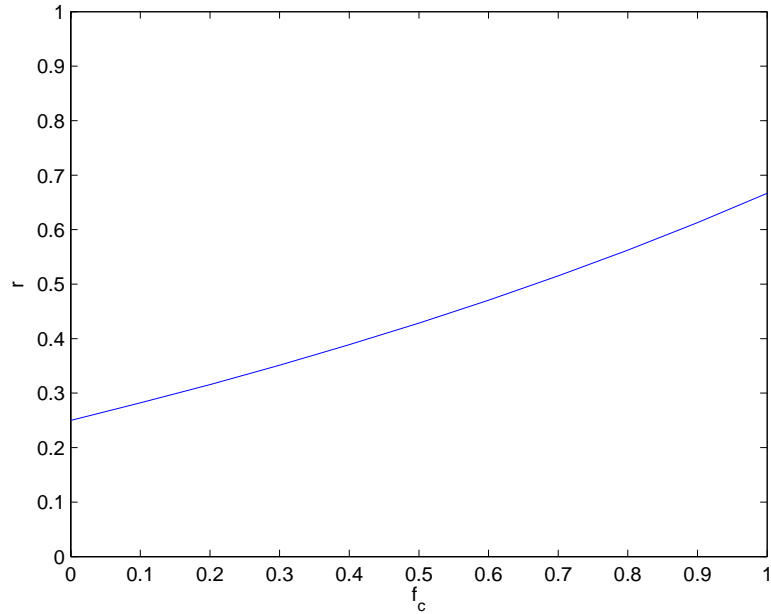

Figure 3. The Relatedness Threshold for Kin-Selected Altruism at Two Loci to be Favoured with Negatively Non-Linear Payoffs

As the population frequency $f_c$ of cooperative alleles at a related locus increases, the degree of relatedness $r'$ required for kin selection to favour altruistic (cooperative) alleles in the Prisoner's Dilemma increases ($T = 5, R = 3, P = 1, S = 0$)

Note that the interval over which the relatedness threshold varies under kin selection involving two loci differs from that under kin selection at a single locus. Consideration of equation (12) shows that $r'$ varies in the closed interval between

$$\frac{P-S}{T-P} \tag{13}$$

and

$$\frac{T-R}{R-S} \tag{14}$$

(c.f. (7) and (8)). If the deviation from additivity $d = 0$, or the frequency of cooperators $f_c = 0$, then the relatedness threshold is given by (13), which is that predicted by Hamilton's rule given the definitions of cost and benefit we have been using. If $d$ is non-zero then which of (13) and (14) is greater depends on its sign.

As for the single locus case, given equations (10) and (11) we can rearrange the condition $i_c = i_d$ to derive an equilibrium frequency of cooperators

$$f'_c = \frac{P-S-r(P-T)}{(1+r)(R-T+P-S)}. \tag{15}$$

Here the equilibrium predicted is the frequency of cooperative behaviour at a locus *other than that at which inclusive fitness is being calculated* ($i_c$ and $i_d$). This is in contrast to equation 9 which predicts the equilibrium for a single locus. Note also that the equilibrium predicted in equation 15 is different to that derived for the single locus case (equation 9). Intriguingly, this equilibrium is identical to that derived by Grafen (1979) for a continuous version of the game between relatives, despite being derived for a discrete game in which the roles of potential altruist and potential beneficiary are distinct. Thus the same equilibrium condition holds for the continuous version of a generic 2 x 2 game in which participants may both behave altruistically or not, and the discrete version of the game in which the roles of donor and recipient are separated.

There is a further difference between the situation analysed here, and that considered by Grafen. Grafen's game can be thought of as involving only one locus, whereas our analysis is for selection involving two loci. At this point it worth considering which is the other locus involved in selection, apart from that determining whether a potential altruist does or does not behave altruistically. In short, what is the 'other' locus referred to by $f_c$ in equations (10) and (11)? It is the locus that determines the behaviour of the potential recipient of altruism. Thus the equilibrium given by equation 15 is the equilibrium frequency of cooperative behaviour among *potential recipients of altruism*. If we are modelling selection at the "forgiving / vengeful" locus, $f_c$ will be the frequency of cooperation alleles at the "merciful / exploitative" locus. Additionally, as our 2-dimensional strategies are symmetric for these two loci, if we are modelling selection at the "merciful / exploitative" locus, $f_c$ will be the frequency of cooperation alleles at the "forgiving / vengeful" locus. As selection at one locus will change a key quantity used in determining selection at the other locus, and vice-versa, we may couple the equations for selection at these two loci. To achieve this we instantiate equations (10) and (11) for each of the loci, thus the inclusive fitnesses of cooperation alleles at each of the two loci are

$$i_{c_1} = Rf_{c_2} + S(1 - f_{c_2}) + r(Rf_{c_2} + T(1 - f_{c_2})), \tag{16}$$

and

$$i_{c_2} = Rf_{c_1} + S(1 - f_{c_1}) + r(Rf_{c_1} + T(1 - f_{c_1})). \tag{17}$$

Similarly the inclusive fitnesses of defection alleles at the two loci are given by

$$i_{d_1} = Tf_{c_2} + P(1 - f_{c_2}) + r(Sf_{c_2} + P(1 - f_{c_2})), \tag{18}$$

and

$$i_{d_2} = Tf_{c_1} + P(1 - f_{c_1}) + r(Sf_{c_1} + P(1 - f_{c_1})). \tag{19}$$

Then we can utilise the replicator dynamics (Schuster & Sigmund, 1983) to give an idealised approximation of the evolutionary pressure for change in frequency of cooperative alleles at each locus as

$$\dot{f}_{c_1} = f_{c_1}\left(i_{c_1} - \left(f_{c_1}i_{c_1} + (1 - f_{c_1})i_{d_1}\right)\right), \tag{20}$$

and

$$\dot{f}_{c_2} = f_{c_2}\left(i_{c_2} - \left(f_{c_2}i_{c_2} + (1 - f_{c_2})i_{d_2}\right)\right). \tag{21}$$

Figures 4 and 5 below illustrate this evolutionary pressure for two cases, the Prisoner's Dilemma with negatively non-additive payoffs, and with positively non-additive payoffs, respectively. In both cases, an intermediate level of relatedness $r$ has been chosen that lies within the open interval whose bounds are given by (13) and (14). In both figures, the equilibrium for each locus given by equation (15) is plotted. Note that, in both cases, the fixed point where both loci are at the mixed equilibrium is a saddle point. In the negatively non-additive payoff case (figure 4), selection will tend to fix the population on a combination of cooperation allele at one locus and defection allele at the other. In the positively non-additive payoff case (figure 5), selection will generally result in the population becoming fixated either for cooperation alleles at both loci, or defection alleles at both loci.

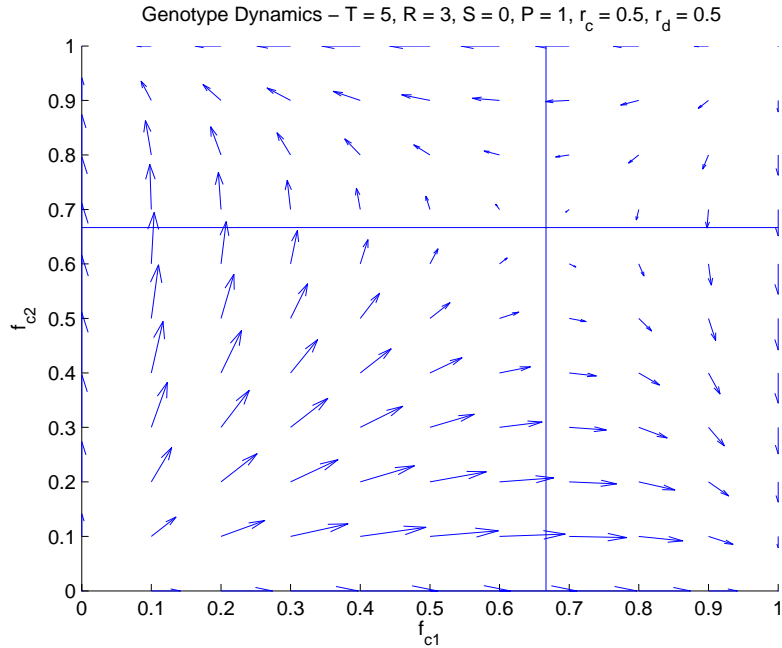

Figure 4. Selection Pressure on "Merciful / Exploitative" and "Forgiving / Vengeful" Loci with Negatively Non-Linear Payoffs
Negative non-linearity in the payoff matrix introduces a saddle point and two stable attractors, corresponding to fixation of the cooperation allele at one of the loci and fixation of the defection allele at the other locus ($T = 5, R = 3, P = 1, S = 0$)

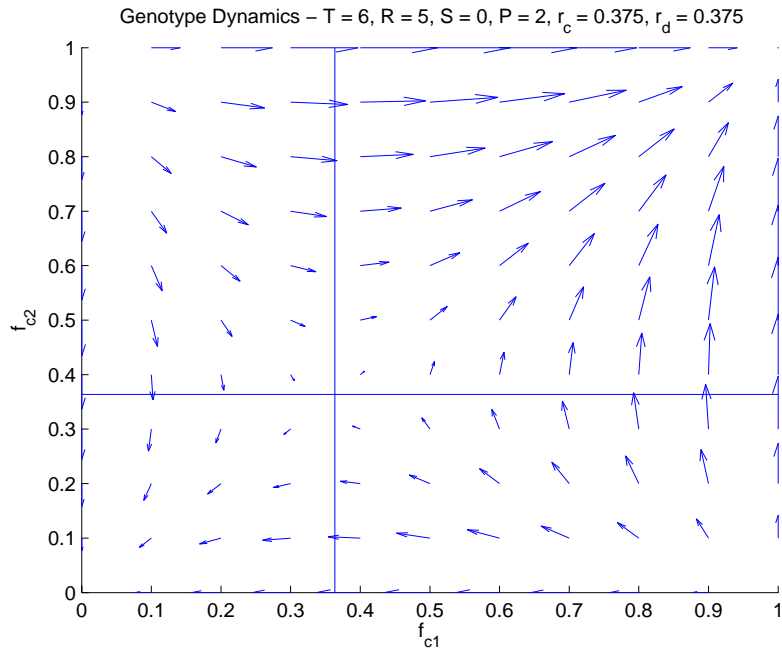

Figure 5. Selection Pressure on "Merciful / Exploitative" and "Forgiving / Vengeful" Loci with Positively Non-Linear Payoffs
Positive non-linearity in the payoff matrix introduces a saddle point and two stable attractors, corresponding to fixation of cooperation alleles at both loci, and fixation of defection alleles both loci ($T = 6$, $R = 5$, $P = 2$, $S = 0$)

**The Hawk-Dove Game**

One further game deserves our consideration. While previous authors have studied a generic 2 x 2 game under the label 'Hawk-Dove', the payoff matrix for this game has been given more specifically by its original proponents (Maynard Smith & Price, 1973). The Hawk-Dove game with averaged payoffs is presented below in table 8.

|   | H | D |
|---|---|---|
| H | $\dfrac{V - C}{2}$ | $V$ |
| D | $0$ | $\dfrac{V}{2}$ |

Table 8. Hawk-Dove Payoff Matrix (Maynard Smith & Price, 1973)
*(note: payoffs to the row player are shown)*

Interestingly, this game can be reduced to two of the different games identified by Rapoport & Guyer (1966). If $C < V$ then Hawk-Dove is equivalent to the ordinal-scale Prisoner's Dilemma, whereas if $C > V$ then Hawk-Dove is equivalent to Chicken with ordinal payoffs. This agrees with previous authors' observations on the relationship between Chicken and the Prisoner's Dilemma (Doebeli & Hauert, 2005). The case where $V = C$, however, does not translate to a 2 x 2 game with an ordinal payoff scale, and so falls outside the class of games studied by Rapoport & Guyer (1966). Furthermore, the Hawk-Dove game as described in table 8 above can only realise a Prisoner's Dilemma with non-additive payoffs (see appendix B for the proof). Thus, while Hawk-Dove may seem a more general game, the Prisoner's Dilemma retains its pre-eminence for our purposes due to its ability to realise the additive donation game.

**Conclusion**

Game theory has been and is widely used to study both kin selection and reciprocity as explanations for cooperation. The Prisoner's Dilemma in particular has proved especially popular for such studies. This paper has shown how this popularity is justified because, of the 2 x 2 games, the Prisoner's Dilemma provides the simplest way to model both reciprocal cooperation and kin-selected altruism. However as has been identified here, the Prisoner's Dilemma is also subject to a significant caveat in its application. Care must be taken when choosing the Prisoner's Dilemma payoff matrix as it is easy to deviate from additivity of payoffs, either positively or negatively, while satisfying the usual constraints on the Prisoner's Dilemma, namely $T > R > P > S$ and $T + S < 2R$. Such a deviation from additivity will have different effects according to whether it is positive or negative, and according to whether the roles of potential altruist and potential beneficiary are or are not separated.

In the case where both game participants simultaneously decide whether or not to behave altruistically, negatively non-additive payoffs such as the canonical $T = 5$, $R = 3$, $P = 1$ and $S = 0$, result in altruism being harder to select than Hamilton's rule predicts. In this case, the more frequent altruism becomes in the population, the higher the relatedness threshold becomes for altruism to be favoured. For intermediate levels of relatedness in a population, this results in convergence on a mixed equilibrium for the population where both cooperation and defection alleles are maintained, as previously demonstrated (Grafen, 1979; Queller, 1984). For cases where the roles of potential altruist and beneficiary are separated, we have shown here that negatively non-additive payoffs will tend to result in the fixation of a cooperation allele for individuals in one of the roles, and a defection allele for individuals in the other role. The analysis presented here explains an earlier result on the potential for kin selection to undermine selection for retaliation in the Iterated Prisoner's Dilemma (Marshall & Rowe, 2003a).

In contrast, positively non-additive payoffs such as $T = 6$, $R = 5$, $P = 2$ and $S = 0$, result in altruism being easier to select than predicted by Hamilton's rule, again for the case where both game participants simultaneously decide whether or not to behave altruistically. In this case, the more frequent altruism becomes in the population, the lower the relatedness threshold becomes for altruism to be favoured. For intermediate levels of relatedness in a population, this results in an unstable mixed equilibrium for the population, resulting in fixation of either cooperation or defection alleles, as previously demonstrated (Grafen, 1979; Queller, 1984). For cases where the roles of potential altruist and beneficiary are separated, we have shown that positively non-additive payoffs will tend to result in the fixation of either cooperation or defection alleles for both roles.

The analysis presented in this paper has focussed on discrete choice strategies. Intriguingly, this analysis has unified the analysis of continuous choice strategies with that for discrete choice games in which the role of potential altruist and beneficiary are separated, showing that the game solution is the same in both cases, albeit with different consequences.

Given the obvious interest in investigations into kin selection and reciprocal cooperation, as well as the further interest in their joint operation due to the interactions between them (Trivers, 1971; Axelrod & Hamilton, 1981), it is hoped that the results presented in this paper will help guide future research on those topics. For modelling of reciprocal cooperation, we recommend the continued use of the Prisoner's Dilemma as the social interaction model, but with the caveat that non-additive payoffs and some form of population structure may allow confounding kin selection effects within any such model. In general, therefore, we recommend that in any model of cooperation where kin selection is being studied or may play a role (for example in viscous population models such as Marshall & Rowe (2003b)), the payoff for any game used should be analysed by mapping onto Queller's (1984) synergistic donation game as shown in table (7). This will make explicit the relative magnitude of the costs and benefits of altruism, their sign, and any synergistic deviation from additivity inherent in the game. This recommendation coincides with a recently proposed framework for the modelling of cooperation in social dilemmas (Hauert *et al.*, 2006).

**Acknowledgements**

I thank R. Planqué, P. Sinha-Ray and I. P. Wright for comments on earlier drafts of this paper.

**Appendix A**

This appendix demonstrates that a Prisoner's Dilemma interaction in which opponents use different loci to determine their game choice is equivalent to a donation game in which the roles of potential altruist and potential beneficiary are separated, as described by Queller (1984).

Such an interaction involving different loci in both players occurs when players use a two-dimensional strategy (Mar & St Denis, 1994), and chose opposite actions on their previous interaction. In this case, one player will use the 'forgiving / vengeful' locus to specify their next game choice, while the other uses the 'merciful / exploitative' locus. The equations for inclusive fitness of cooperation and defection alleles at these loci are given by equations 10 and 11 in the main text (from Marshall & Rowe (2000, 2003)). With the definitions of cost $c = P - S$, benefit $b = T - P$ and deviation from additivity $d = R - S - T + P$ used in the text, these equations can be rewritten as

$$i_c = r(b + qd) - c + q(b + d), \tag{A1}$$

and

$$i_d = qb. \tag{A2}$$

From this rewriting, the game in which the 'forgiving / vengeful' and 'merciful / exploitative' loci are involved can be understood as a donation game in which the role of potential altruist and potential beneficiary are separated, but in which there is some additional, possibly synergistic behaviour of the potential recipient. If this synergistic effect $d = 0$, then the frequency with which potential recipients perform this behaviour drops out when we solve for $i_c = i_d$, and the game exactly maps onto Queller's (1984) situation in which the role of potential altruist and potential recipient is separated. In this case the condition for selection for altruism is given solely by the relatedness of the potential donor to the recipient, and the ratio of cost to benefit, as stated in Hamilton's rule (equation 1).

If, however, the synergistic effect $d$ is non-zero we have a synergistic, discrete-choice donation game with roles. Solving for the equilibrium frequency of this synergistic behaviour among potential recipients of altruism, we find we have the same equilibrium as predicted by Grafen (1979) for continuous games between relatives (equation 15). Note that, if $d = 0$, Grafen's equilibrium condition is undefined.

**Appendix B**

This appendix contains proofs of the statements given in this paper on the suitability or unsuitability of the different symmetric non-trivial 2 x 2 games for modelling kin selected altruism.

*Proposition 1.* 'Chicken' cannot realise the donation game with strong altruism.

*Proof.* There are two ways in which we may try to map the Chicken payoff (table 5) matrix onto the donation game payoff matrix (table 1); either we assign choice A in Chicken to be the altruist type, or we assign choice B as altruist. Under either assignment, the payoff corresponding to $-c$ (received when playing A against B, or B against A, respectively) is lower than the payoff corresponding to the baseline fitness when two non-altruists interact (received when playing B against B, or A against A, respectively). Considering baseline fitness as 0, $-c$ can thus be seen to be positive, hence there is no cost for altruism in the game and strong altruism cannot occur. ∎

*Proposition 2.* 'Apology' and 'Battle of the Sexes' cannot realise the donation game with strong altruism.

*Proof.* The proof is the same as for Chicken. ∎

*Theorem 1.* 'Prisoner's Dilemma' is the only symmetric 2 x 2 game that can realise the additive donation game with strong altruism.

*Proof.* First we show that the Prisoner's Dilemma with ordinal payoffs can realise the donation game with strong altruism. The ordinal-payoff Prisoner's Dilemma is as follows (payoffs to row-player shown):

|   | A | B |
|---|---|---|
| A | 1 | -2 |
| B | 2 | -1 |

From this payoff matrix we can see that if A is taken as the altruist type, then the payoff corresponding to $-c$ (that received by playing A against B) is the lowest in the payoff matrix, lower than the baseline fitness received by two non-altruists interacting together. Thus $-c$ has a positive fitness value and there is a genuine fitness cost for the altruist type. Similarly the payoff corresponding to $b$ (that received by playing B against A) is the highest in the payoff matrix, indicating that $b$ has a positive fitness value and there is a genuine fitness benefit for interacting with the altruist type. Finally, the payoff corresponding to $b - c$ (that received by playing A against A) is greater than that corresponding to $-c$ and less than that corresponding to $b$, hence the inequality for the donation game $b > c > 0$ is satisfied.

For the additive donation game, the payoff to an altruist interacting with an altruist should equal the sum of the payoff to an altruist interacting with a non-altruist, and a non-altruist interacting with an altruist. It can easily be seen that a cardinal-payoff Prisoner's Dilemma matrix can be constructed with this property by ensuring $R - S + P - T = 0$, while still satisfying the Prisoner's Dilemma constraints $T > R > P > S$ and $T + S < 2R$, for example (with payoffs to row-player shown)

|   | A | B |
|---|---|---|
| A | 4 | 0 |
| B | 6 | 2 |

The uniqueness of the Prisoner's Dilemma in being able to realise the donation game as described above lies in its non-equivalence with any other symmetric 2 x 2 game (Rapoport & Guyer, 1966) ∎

*Theorem 2.* 'Hawk-Dove' with averaged payoffs cannot be parameterised for equivalence with an additive-payoff 'Prisoner's Dilemma'.

*Proof.* Regardless of whether we select D or H to be the altruistic type in the Hawk-Dove game (table 8), from the additive donation game (table 1) the difference between the payoffs available to the row player should be $c$ irrespective of the column player's choice. Considering the payoffs in the first column of the Hawk-Dove payoff matrix this indicates $c = \dfrac{V}{2}$, while from the second column we should find $c = \dfrac{V - C}{2}$. To satisfy both these equalities we must have $C = 0$, in which case it can be seen that the payoff to two altruist interacting together is the same as the payoff to two non-altruists, and the game is no longer in the class of ordinal-payoff 2 x 2 games to which the Prisoner's Dilemma belongs. ∎